\documentclass[twocolumn]{el-author}

\usepackage{epsfig}
\usepackage{booktabs}
\usepackage{epstopdf}

\newcommand{\ua}{\uparrow}
\newcommand{\nc}{\newcommand}
\nc{\da}{\downarrow} \nc{\hc}{\hat{c}} \nc{\hS}{\hat{S}}
\nc{\bra}{\langle} \nc{\ket}{\rangle} \nc{\eq}{equation (\ref}
\nc{\h}{\hat} \nc{\hT}{\h{T}}\nc{\be}{\begin{eqnarray}}
\nc{\ee}{\end{eqnarray}}\nc{\rd}{\textrm{d}}\nc{\e}{eqnarray}\nc{\hR}{\hat{R}}\nc{\Tr}{\mathrm{Tr}}
\nc{\tS}{\tilde{S}}\nc{\tr}{\mathrm{tr}}\nc{\8}{\infty}\nc{\lgs}{\bra\ua,\phi|}\nc{\rgs}{|\ua,\phi\ket}
\nc{\hU}{\hat{U}}\nc{\lfs}{\bra\phi|}\nc{\rfs}{|\phi\ket}\nc{\hZ}{\hat{Z}}\nc{\hd}{\hat{d}}\nc{\mD}{\mathcal{D}}
\nc{\bd}{\bar{d}}\nc{\bc}{\bar{c}}\nc{\mc}{\mathcal}\nc{\ea}{eqnarray}\nc{\mG}{\mathcal{G}}\nc{\bce}{\begin{center}}
\nc{\ece}{\end{center}}
\date{12th December 2011}

\begin{document}

\title{Secrecy Outage Analysis Over Fluctuating Two-Ray Fading Channels}

\author{Hui Zhao, Liang Yang,  Gaofeng Pan,  and M.-S. Alouini}

\abstract{In this letter, we analyze the secrecy outage probability (SOP) over fluctuating two-ray fading channels but with a different definition from the one adopted in \cite{Zhang2}. Following the new defined SOP, we derive an analytical closed-form expression for our proposed SOP, as well as an asymptotic formula valid in the high signal-to-noise ratio region of the source to destination link. In the numerical results section, we perform some Monte-Carlo simulations to validate the accuracy of our derived expressions, and also present the probability gap between our proposed SOP and the SOP in \cite{Zhang2}.}

\maketitle

\section{Introduction}
Although Rayleigh and Rician distributions can be used in modeling the fading channels in millimeter-wave (mmWave) bands, the random fluctuation suffered by the received signal cannot be analyzed in those conventional fading models. To capture mmWave fading channels more accurately, \cite{Goldsmith} proposed a new channel model, called the fluctuating two-ray (FTR) model, which provides a much better fit than Rician model in 28 GHz outdoor mmWave channels. This FTR channel is also a natural generalization of the two-wave with diffuse power (TWDP) model in \cite{Durgin}. Compared to the TWDP model, the amplitude of specular waves randomly fluctuates in the FTR model. Moreover, the FTR model can be reduced to many conventional fading models, such as one-sided Gaussian, Nakagami-$m$, and Rcian shadowed fading models. However, the FTR model in \cite{Goldsmith} is very complicated and difficult for performance analysis. To address this issue, \cite{Zhang} modified and extended the FTR model in \cite{Goldsmith}, which provides a relatively simple expression and allows $m$, a parameter of FTR model, to take any positive value, rather than only integers allowed in \cite{Goldsmith}.

More recently, the capacity under different power adaption schemes over FTR channels has been analyzed in \cite{Hui} based on the work of \cite{Zhang}. The authors of \cite{Zhang2} investigated the physical layer security over FTR channels, and derived the closed-form expression for secrecy outage probability (SOP) under the traditional SOP definition in \cite{Bloch}. This SOP definition has been adopted in the majority of physical layer security works, such as  \cite{Pan_TCOM}-\cite{Zhao_FITEE}.                       However,
 this SOP definition in \cite{Bloch} cannot give the exact probability of information leakage to the eavesdropper, because unreliable transmission from the source to the destination is also considered as secrecy outage. \cite{Zhou} proposed an alternative SOP formulation, where the information leakage is the only case of secrecy outage, and the transmitter can adjust the reliable transmission threshold to compromise the security and quality of service of the legitimate receiver.

 In this letter, we adopt the SOP definition in \cite{Zhou} to analyze the secrecy outage performance, and derive a closed-form expression for the SOP. To simplify the expression for SOP and get some insights, we then also investigate the asymptotic SOP (ASOP) in the high signal-to-noise ratio (SNR) region of the source to destination link. Finally, we compare the probability gap between two different SOP definitions in \cite{Bloch,Zhou}.

\section{System Model}
There is a source ($S$) transmitting confidential message to a destination ($D$), while an eavesdropper ($E$) wants to overhear the information from $S$ to $D$. We assume that all links undergo independent FTR fading channels. Let $\gamma_d$ and $\gamma_e$ be the instantaneous SNRs at $D$ and $E$, respectively.
The probability density function (PDF) and cumulative density function (CDF) of $\gamma_d$ and $\gamma_e$ over FTR fading channels are \cite{Zhang,Hui}
\begin{align}
&{f_{{\gamma _t}}}\left( x \right) = \frac{{m_t^{{m_t}}}}{{\Gamma \left( {{m_t}} \right)}}\sum\limits_{{j_t} = 0}^\infty  {\frac{{K_t^{{j_t}}{d_{{j_t}}}{x^{{j_t}}}\exp \left( { - \frac{x}{{2\sigma _t^2}}} \right)}}{{{j_t}!{j_t}!{{\left( {2\sigma _t^2} \right)}^{{j_t} + 1}}}}}, \\
&{F_{{\gamma _t}}}\left( x \right) = 1 - \frac{{m_t^{{m_t}}}}{{\Gamma \left( {{m_t}} \right)}}\sum\limits_{{j_t} = 0}^\infty  {\frac{{K_t^{{j_t}}{d_{{j_t}}}}}{{{j_t}!\exp \left( {\frac{x}{{2\sigma _t^2}}} \right)}}} \sum\limits_{{n_t} = 0}^{{j_t}} {\frac{{{{\left( {{x \mathord{\left/
 {\vphantom {x {\left( {2\sigma _t^2} \right)}}} \right.
 \kern-\nulldelimiterspace} {\left( {2\sigma _t^2} \right)}}} \right)}^{{n_t}}}}}{{{n_t}!}}},
\end{align}
respectively, where $t \in \{d,e\}$, $m_t$ is the parameter of Gamma distribution with unit mean, $\sigma_t^2$ denotes the variance of the real (or imaginary) part of diffuse components, $K_t$ is the average power ratio of the dominant waves to the remaining diffuse multipath, and $d_{j_d}$ is given by
\begin{align}
{d_{{j_t}}} =& \sum\limits_{k = 0}^{{j_t}} {j_t \choose k} {\left( {\frac{{{\Delta _t}}}{2}} \right)^k}\sum\limits_{l = 0}^k {k \choose l}\Gamma \left( {{j_t} + {m_t} + 2l - k} \right) \notag\\
&\exp \left( {\frac{{i\pi \left( {2l - k} \right)}}{2}} \right)\frac{{P_{{j_t} + {m_t} - 1}^{k - 2l}\left( {\frac{{{m_t} + {K_t}}}{{\sqrt {{{\left( {{m_t} + {K_t}} \right)}^2} - {{\left( {{K_t}{\Delta _t}} \right)}^2}} }}} \right)}}{{{{\left[ {\sqrt {{{\left( {{m_t} + {K_t}} \right)}^2} - {{\left( {{K_t}{\Delta _t}} \right)}^2}} } \right]}^{{j_t} + {m_t}}}}},
\end{align}
where ${{\Delta _t}} \in [0,1]$ is a parameter to show the similarity of the average received powers of different specular components, $i$ denotes the imaginary unit, $P(\cdot)$ and $\Gamma(\cdot)$ denote Legendre functions of the first kind and Gamma function \cite{Gradshteyn}, respectively.

From the expression for the CDF of $\gamma_t$, it is obvious that the complementary CDF (CCDF) of $\gamma_t$ is
\begin{align}
{\overline F _{{\gamma _t}}}\left( x \right) = \frac{{m_t^{{m_t}}}}{{\Gamma \left( {{m_t}} \right)}}\sum\limits_{{j_t} = 0}^\infty  {\frac{{K_t^{{j_t}}{d_{{j_t}}}}}{{{j_t}!\exp \left( {\frac{x}{{2\sigma _t^2}}} \right)}}} \sum\limits_{{n_t} = 0}^{{j_t}} {\frac{{{{\left( {{x \mathord{\left/
 {\vphantom {x {\left( {2\sigma _t^2} \right)}}} \right.
 \kern-\nulldelimiterspace} {\left( {2\sigma _t^2} \right)}}} \right)}^{{n_t}}}}}{{{n_t}!}}}.
\end{align}

\section{Secrecy Outage Probability}
We assume that $S$ does not know the instantaneous channel state information of  $S-E$ link, i.e., silent eavesdropping. In this case, perfect security cannot be guaranteed, because $S$ has no choice but to adopt the constant rate of confidential information ($R_s$). SOP is used to evaluate the secrecy outage performance, which is the probability that the instantaneous secrecy capacity ($C_s$) is less than $R_s$ in \cite{Bloch}, where $C_s$ is defined as
$
{C_s} = \max \left\{ {{{\log }_2}\left( {1 + {\gamma _d}} \right) - {{\log }_2}\left( {1 + {\gamma _e}} \right),0} \right\},
$
where $C_d={\log _2}\left( {1 + {\gamma _d}} \right)$ and $C_e={\log _2}\left( {1 + {\gamma _e}} \right)$ are the capacity of $S-D$ and $S-E$ links,  respectively.
Therefore, the SOP in \cite{Bloch} can be written as
$
{\rm{SOP}} =  \Pr \left\{ {{{\log }_2}\left( {1 + {\gamma _d}} \right) - {{\log }_2}\left( {1 + {\gamma _e}} \right) \le {R_s}} \right\},
$
In this conventional SOP definition,  both the unreliable transmission from $S$ to $D$ and the information leakage to $E$ are considered as secrecy outage, which means that this secrecy outage cannot necessarily imply a  perfect secure failure. To capture the actual information leakage probability, \cite{Zhou} proposed an alternative SOP definition, which is called the modified SOP in this letter, and which is given by
\begin{align}
{\rm{SOP}} = \Pr \left\{ {{C_e} > {C_d} - {R_s}\left| {{\gamma _d} > \mu } \right.} \right\}
 = \frac{{\Pr \left\{ {\mu  < {\gamma _d} < \lambda  - 1 + \lambda {\gamma _e}} \right\}}}{{\Pr \left\{ {{\gamma _d} > \mu } \right\}}},
\end{align}
where $\lambda=2^{R_s}$. In the modified SOP formulation, if the received SNR at $D$ is greater than a certain threshold ($\mu$), $S$ transmits signals to $D$, because $D$ is able to decode the signal from $S$, i.e., a reliable transmission. If this reliable transmission happens, $E$ has the opportunity to wiretap the information from $S$ to $D$, otherwise no possibility of information leakage to $E$.

Using some probability theory knowledge, we can easily rewrite the modified SOP as
\begin{align}\label{SOP_int}
&{\rm{SOP}} = \frac{{\int_{\frac{{\mu  + 1}}{\lambda } - 1}^\infty  {\int_\mu ^{\lambda  - 1 + \lambda {\gamma _e}} {{f_{{\gamma _d}}}\left( x \right)} } dx{f_{{\gamma _e}}}\left( y \right)dy}}{{{{\overline F }_{{\gamma _d}}}\left( \mu  \right)}} \notag\\
& = \frac{{\int_{\frac{{\mu  + 1}}{\lambda } - 1}^\infty  {{F_{{\gamma _d}}}\left( {\lambda  - 1 + \lambda y} \right)} {f_{{\gamma _e}}}\left( y \right)dy}}{{{{\overline F }_{{\gamma _d}}}\left( \mu  \right)}}
 - \frac{{ {F_{{\gamma _d}}}\left( \mu  \right){{\overline F }_{{\gamma _e}}}\left( {\frac{{\mu  + 1}}{\lambda } - 1} \right)}}{{{{\overline F }_{{\gamma _d}}}\left( \mu  \right)}},
\end{align}
By substituting the PDF of $\gamma_e$ and the CDF of $\gamma_d$ into this modified SOP definition and using some mathematical manipulations, 
the SOP over FTR fading channels can be derived as
\begin{align}\label{SOP}
{\rm{SOP}} =& {\overline F _{{\gamma _e}}}\left( {\frac{{\mu  + 1}}{\lambda } - 1} \right) - \frac{{m_d^{{m_d}}}}{{\Gamma \left( {{m_d}} \right){{\overline F }_{{\gamma _d}}}\left( \mu  \right)}}\sum\limits_{{j_d} = 0}^\infty  {\frac{{K_d^{{j_d}}{d_{{j_d}}}}}{{{j_d}!}}}  \notag\\
&\sum\limits_{{n_d} = 0}^{{j_d}} {\frac{{\exp \left( { - \frac{{\lambda  - 1}}{{2\sigma _d^2}}} \right)}}{{{n_d}!{{\left( {2\sigma _d^2} \right)}^{{n_d}}}}}} \sum\limits_{k = 0}^{{n_d}} {n_d \choose k} {\left( {\lambda  - 1} \right)^{{n_d} - k}}{\lambda ^k}\frac{{m_e^{{m_e}}}}{{\Gamma \left( {{m_e}} \right)}} \notag\\
&\sum\limits_{{j_e} = 0}^\infty  {\frac{{K_e^{{j_e}}{d_{{j_e}}}}}{{{j_e}!{j_e}!{{\left( {2\sigma _e^2} \right)}^{{j_e} + 1}}}}} \frac{{\Gamma \left( {{j_e} + k + 1,\left( {\frac{\lambda }{{2\sigma _d^2}} + \frac{1}{{2\sigma _e^2}}} \right)\left( {\frac{{\mu  + 1}}{\lambda } - 1} \right)} \right)}}{{{{\left( {\frac{\lambda }{{2\sigma _d^2}} + \frac{1}{{2\sigma _e^2}}} \right)}^{{j_e} + k + 1}}}},
\end{align}
where $\Gamma(\cdot,\cdot)$ denote the upper incomplete Gamma function \cite{Gradshteyn}.

\section{Asymptotic Analysis}
To simplify the expression for the modified SOP and drive the secrecy diversity order, we perform the asymptotic analysis for the modified SOP when the average SNR of $S-D$ link goes to infinity, i.e., $\overline \gamma_d \to \infty$.

From Eq. (5) in \cite{Zhang}, the relationship between $\overline \gamma_d$ and $2\sigma_d^2$ is given by
$
{\overline \gamma  _d} = \frac{{{P_t}}}{{{N_0}}}\left( {1 + {K_d}} \right)2\sigma _d^2,
$
where $P_t$ and $N_0$ are the transmit power at $S$ and the Gaussian noise power at $D$, respectively. This relationship shows that  $\overline \gamma_d \to \infty$ implies $2\sigma_d^2 \to \infty$.
For $2\sigma_d^2 \to \infty$, the asymptotic CDF of $\gamma_d$ is \cite{Goldsmith,Zhang}
\begin{align}
F_{{\gamma _d}}^\infty \left( x \right) = \frac{{m_d^{{m_d}}{d_{0}}x}}{{\Gamma \left( {{m_d}} \right)}}{\left( {2\sigma _d^2} \right)^{ - 1}} + o\left( {{{\left( {2\sigma _d^2} \right)}^{ - 2}}} \right),
\end{align}
where $o(\cdot)$ denotes the higher order term, and $d_{0}$ is the value of $d_{j_d}$ when $j_d=0$, given by
\begin{align}
{d_0} = \frac{{\Gamma \left( {{m_d}} \right){P_{{m_d} - 1}}\left( {\frac{{{m_d} + {K_d}}}{{\sqrt {{{\left( {{m_d} + {K_d}} \right)}^2} - {{\left( {{K_d}{\Delta _d}} \right)}^2}} }}} \right)}}{{{{\left[ {\sqrt {{{\left( {{m_d} + {K_d}} \right)}^2} - {{\left( {{K_d}{\Delta _d}} \right)}^2}} } \right]}^{{m_d}}}}}.
\end{align}
When $2\sigma_d^2 \to \infty$, the integral form for the modified SOP in \eqref{SOP_int} can be approximated as
\begin{align}
{\rm{SO}}{{\rm{P}}^\infty } \approx & \int_{\frac{{\mu  + 1}}{\lambda } - 1}^\infty  {F_{{\gamma _d}}^\infty \left( {\lambda  - 1 + \lambda y} \right)} {f_{{\gamma _e}}}\left( y \right)dy \notag\\
& - F_{{\gamma _b}}^\infty \left( \mu  \right){\overline F _{{\gamma _e}}}\left( {\frac{{\mu  + 1}}{\lambda } - 1} \right).
\end{align}
After some mathematical manipulations, ASOP can be derived as
\begin{align}
&{\rm{SO}}{{\rm{P}}^\infty } = {\left( {2\sigma _d^2} \right)^{ - 1}}\left\{ {\frac{{m_d^{{m_d}}{d_0}\left( {\lambda  - 1 - \mu } \right)}}{{\Gamma \left( {{m_d}} \right)}}{{\overline F }_{{\gamma _e}}}\left( {\frac{{\mu  + 1}}{\lambda } - 1} \right) + } \right. \notag\\
&\left. {\frac{{m_d^{{m_d}}m_e^{{m_e}}{d_0}\lambda 2\sigma _e^2}}{{\Gamma \left( {{m_d}} \right)\Gamma \left( {{m_e}} \right)}}\sum\limits_{{j_e} = 0}^\infty  {\frac{{K_e^{{j_e}}{d_{{j_e}}}\Gamma \left( {{j_e} + 2,\left( {\frac{{\mu  + 1}}{\lambda } - 1} \right)2\sigma _e^2} \right)}}{{{j_e}!{j_e}!}}} } \right\},
\end{align}
which shows that the secrecy diversity order is always unity. Note that although Nakagami-$m$ fading model is a limit case of the FTR model, some mathematical properties change in that limit case, which will offer different conclusions for the secrecy diversity order. Due to the page limitation, we do not consider that limit case.

\section{Numerical Results}
To simplify the parameter setting, $P_t/N_0=0$ dB, $m_d=m_e=m$, $K_d=K_e=K$, $\Delta_d=\Delta_e=0.5$, $R_s=1$ are assumed in the following simulation results. In the calculation of the analytical results, we truncate the infinite summation terms into finite terms, where the truncated error has been analyzed in \cite{Zhang,Zhang2}.
\begin{figure}[!htb]
\setlength{\abovecaptionskip}{0pt}
\setlength{\belowcaptionskip}{10pt}
\centering
\includegraphics[width= 2.5 in]{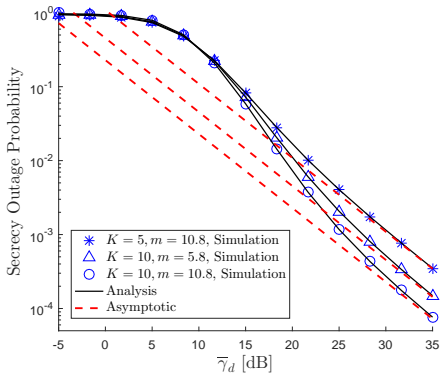}
\caption{SOP versus $\overline \gamma_d$ for $\overline \gamma_e=5$ dB and $\mu=2$.}\vspace{-0.5cm}
\label{P_D_muH}
\end{figure}

Fig. 1 plots the modified SOP versus $\overline \gamma_d$ for different $m$ and $K$, where the improving trend of SOP is obvious with increasing $\overline \gamma_d$, due to the improved quality of main channel. The increase of $K$ and $m$ means more stronger dominant waves, leading to a better SOP. In the high SNR region, the derived asymptotic results match the exact results very well, where the fixed slope means the unit diversity order.

From Fig. 2, we can easily see that the difference between the conventional SOP in \cite{Zhang2,Bloch} and the modified SOP converges as $\mu$ decreases, because the decline in $\mu$ implies more transmission in worse channel states. Besides, the SOP becomes large as $\overline \gamma_e$ improves, because of the improved wiretap channel.

\begin{figure}[!htb]
\vspace{-0.3cm}
\setlength{\abovecaptionskip}{0pt}
\setlength{\belowcaptionskip}{10pt}
\centering
\includegraphics[width= 2.5 in]{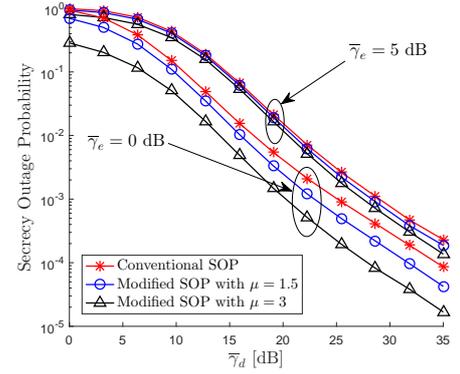}
\caption{SOP versus $\overline \gamma_d$ for $m=3.5$ and $K=15$.}\vspace{-1cm}
\label{P_D_muH}
\end{figure}

\section{Conclusion}
A closed-form expression for the modified SOP over FTR fading channels was derived. An asymptotic formula valid in high SNRs, which reveals the unit secrecy diversity order, was also presented. The comparison between the conventional SOP and modified SOP was discussed in the simulation section, showing a convergence of those two SOP results as $\mu$ decreases.


\vskip5pt

\noindent H. Zhao (\textit{Computer, Electrical, and Mathematical Science
and Engineering  Division, King Abdullah University of Science and Technology (KAUST), Thuwal  23955-6900, Saudi Arabia}, and \textit{Communication Systems Department, EURECOM, Sophia Antipolis 06410, France. The work of H. Zhao was done while he was working at KAUST.})
\vskip5pt
\noindent L. Yang (\textit{College of Computer Science and Electronic
Engineering, Hunan University, Changsha 410082, China.})
\vskip5pt
\noindent G. Pan (\textit{School of Information and Electronics, Beijing Institute of Technology, Beijing 100081, China.})
\vskip3pt
\noindent E-mail: penngaofeng@qq.com
\vskip5pt
\noindent M.-S. Alouini (\textit{Computer, Electrical, and Mathematical Science
and Engineering  Division, King Abdullah University of Science and Technology, Thuwal  23955-6900, Saudi Arabia})

\end{document}